\documentstyle[a4]{article}
\begin{document}

\newcommand{\be}{\begin{equation}}
\newcommand{\ee}{\end{equation}}
\newcommand{\epe}{\end{equation}}
\newcommand{\bea}{\begin{eqnarray}}
\newcommand{\eea}{\end{eqnarray}}
\newcommand{\ba}{\begin{eqnarray*}}
\newcommand{\ea}{\end{eqnarray*}}
\newcommand{\epa}{\end{eqnarray*}}
\newcommand{\ar}{\rightarrow}

\def\r{\rho}
\def\D{\Delta}
\def\R{I\!\!R}
\def\l{\lambda}
\def\D{\Delta}
\def\d{\delta}
\def\T{\tilde{T}}
\def\k{\kappa}
\def\t{\tau}
\def\f{\phi}
\def\p{\psi}
\def\z{\zeta}
\def\ep{\epsilon}
\def\hx{\widehat{\xi}}
\def\a{\alpha}

\begin{center}

{\large On the Microscopical Structure of the Classical
Spacetime.}

\vspace{0.6cm}

Marcelo Botta Cantcheff\footnote{botta@cbpf.br}

\vspace{3mm}

 Centro Brasileiro de Pesquisas Fisicas (CBPF)

Departamento de Teoria de Campos e Particulas (DCP)

Rua Dr. Xavier Sigaud, 150 - Urca

22290-180 - Rio de Janeiro - RJ - Brazil.

\end{center}

\begin{abstract}

Our purpose here is to introduce the idea of viewing the spacetime
as a macroscopic
 complex system which, consequently, cannot be directly quantized. It should be thought of as
  a collection of more fundamental "microscopical" entities (atoms of geometry), much
   like a solid system,
 in which an atomic (classical) structure must be first recognized in order
 to ensure a correct and meaningful quantization procedure. In other words, we claim
  that the classical limit from a quantum theory of gravity
could not give a four dimensional Einstein spacetime directly, but
requiring {\it a further} macroscopical limit.
 This is analogous to a material medium, whose complete
 description does not come from any quantized field.

We also discuss a possible realization of this hypothesis for
black hole spacetimes.

\end{abstract}
%\end{center}

%\newpage

%\section{Introduction.}

\vspace{0.7cm}

Recognition of the true degrees of freedom associated to the space
time geometry is the central puzzle one has to solve in order to
formulate a theory of quantum gravity.
 It has already been suggested by many authors that
General Relativity could be an effective theory emerging from
another more fundamental one;
 however, these speculations have been conceived and eventually worked out as a low energy limit
 rather than in the sense we explore here, where an eventual microscopic
 structure of the geometry is pointed out.
A straight quantization should not make sense in a such scenario.

 One should notice that it would be impossible to find the molecular
structure of a fluid by directly quantizing some field. To obtain
a correct quantum description of a matter system (gas, solid or
liquid) one
 needs to recognize previously the structure of its microscopical components.

%and taking the macroscopical limit

In this note we adopt a similar point of view for the spacetime
geometry \cite{pad}. To do that, we propose the following general
hypothesis:

\vspace{0.4cm}

I. {\it There is a microscopical geometrical structure (M.G.S, the
space time to small scale) which approaches a smooth 4-dimensional
manifold with a Lorentzian metric (given by the Einstein theory)
$(M,g_{ab})$ in the proper thermodynamical/macroscopical
 limit.}

\vspace{0.4cm}

Let us explain this more precisely. As we state, the concept of
thermodynamic limit is similar to the one adopted for matter
systems: one must average over many microscopical components
(sub-geometries) distributed over macroscopical distances
\cite{pad}. In principle, this concept is not independent of the
strength of the gravitational field, since its magnitude should
give a scale where the macroscopical geometry is recovered.
Notice, in addition, that this is not a classical limit, and the
microscopical scale needs not be related to the Planck's scale a
priori.

This is a statement about the {\it classical} structure of
spacetime rather than on its quantization.
 It does not encode any information about quantum gravity, but
 refers to the true {\it classical} degrees of freedom of the spacetime as a starting
  point for an eventual quantization.
This hypothesis describes a perspective with respect to which the
spacetime should be classically conceived, and is independent of
the particular model for the specific microscopical structure.

One can speculate on the nature of this MGS. The simplest (but
perhaps not unique) point of view is that such micro-geometry
shall constitute the background in itself where all the physical
fields propagate. So, more specifically, its topological structure
should consist in {\it a set ${\cal M}$, which can be decomposed
in the
 union of differentiable manifolds}; roughly speaking, a sort of {\it medium} formed by many
  lower dimensional manifolds \footnote{But we are not assuming any pre-specified dimension for
   these component manifolds.}. One could also admit connectness of this set, in order
    to allow some causal correlation among all its points, and/or existence of a metric on each
     smooth submanifold, in order to have classical propagation of fields. However, this could be too strong.

Two important facts about the quantum gravity problem follow
immediately from (I) (and motivate it):

\begin{itemize}

\item "Microscopic geometries should be suitable to
be quantized ": if the dimension of the micro-manifolds is low,
quantum fields,
 including gravitons, are more tractable. For instance, if the microscopical
  spacetime is a sort of (disjoint) union of smooth manifolds $ M_i $
with $dim M_i < 4 $, then the Hilbert space for quantum gravity
should simplify to $H = \otimes_i H_i$.

\item The often commented mysterious relation between geometry and thermodynamics
 becomes more clear from this viewpoint. It sheds light on the
  statistical interpretation of the black hole thermodynamics and the
   Beckenstein-Hawking entropy, since the spacetime must now be regarded as a
   system of sufficient complexity,
    which behaves ergodically, leading to a coarse grained dissipative description.

\end{itemize}

Next, we present a possible microscopical structure of the space
time geometry based on the existence of horizons, which is
believed to be a generical behavior of spacetime
 by virtue of the censorship conjecture.
 Several aspects of this model bring about (curious) similarity, with recent descriptions of black holes
 in terms of branes, in the context of superstrings and matrix theories, as shall be noted
by the reader. However, we will not discuss this in detail.

{\bf An example of Microscopical Geometry:}

%Black Holes provide a testing ground for quantum gravity

We do not have any strong physical indication about what
 the microscopical picture of the geometry may be, but Black Holes
 should provide us with the proper testing ground due to their
  thermodynamic properties.

This model should be considered to describe the (kinematic)
microscopical
 structure of general spacetimes (with horizons), supposed to be exact
   in the near horizon region where the gravitational field is strong, and deformed at huge
   distances
  from each connected component of the black hole.
    So, for simplicity, let us consider, a static globally hyperbolic spacetime $(M,
g_{\mu \nu})$ with a single black hole region endowed with a
boundary, the event horizon $H$, which is an achronal
hypersurface. Let also be a foliation $\Sigma_t$ of Cauchy
surfaces parameterized by a global time function $t$ and the
spatial horizon $h_t \equiv \Sigma_t \cap H$ , which is a
2-dimensional Riemmanian surface.

We describe now the M.G.S., or more strongly, {\it we define the
space
  time geometry in itself} in such a way that in the thermodynamic limit
   it approaches to $(M, g_{\mu \nu})$:

The (classical) microscopical near horizon space time (detected by
distant fiducial observers)
 consists in a finite, totally ordered set of 3-dimensional, differentiable, homeomorphic to $H$
and non-intersecting manifolds $H_n \,\, n=1,...,N$ (H-branes to
distinguish them from
 ''other" branes in the literature),
all of them equipped with a metric $q_{\mu \nu} \, , \mu \nu =
0,1,2 $ together with a collection of "links" ${\cal J}$ which
consist in smooth manifolds joining them (their boundaries lie on
H-branes).
 Here we
will not describe them in more detail but clearly these links can
only be one, two-dimensional (observe the curious similarity with
open strings and $0$-branes) or even three-dimensional, which
would be a throat between two H-brane.

By taking an injective map $\r: [1, N] \to [0, R] \subset \R$, we
define $H_n \to H_{\r(n)}$ and using that $H_n $ are
 homeomorphic to $H$, we get
a natural map $\r : \bigcup_n H_n \to [0, R] \times H \sim M_R $.
In this sense, a four dimensional base manifold $ M_R \sim [0, R]
\times H $ naturally emerges from the microscopical structure, and
approaches a continuum for a large $N$ and distant observers. We
identify this $M_R$ with the macroscopic near horizon spacetime.
Due to this property, this model turns out to be the simplest one
 to recover the correct dimensionality for the macroscopic
manifold. For other microscopical structures it would be
technically more complicate.

 This is a sort of H-brane gas (or solid) which is not placed in any background
 geometry with fixed metric; instead, it {\it constitutes} the background
 geometry in itself, where only
the embedding field $\r_n: H_n \to M_R $ has physical meaning. No
metric for $M_R$ is assumed, differently from other brane
pictures, so, an effective
 notion of distance
 between branes
should be related to their effective coupling. It should depend on
the fields
 $\r_n$ and other dynamical fields
defined on the H-branes which depend on the links structure ${\cal
J}$.
 As we pointed out before, we expect that this
 effective distance decreases far from the black hole (weak field) and the branes
 topology be deformed. Notice that general covariance would be violated near the horizon,
 but there should be no measurable distortion at the macroscopical/weak field limit
  \footnote{The possible microscopical violation of the Lorentz symmetry
  has been recently explored in the literature \cite{BLS}.}.
This constitutes the kinematical structure of the classical
geometry at microscopical level.

This model resembles the membrane paradigm \cite{libro}, where the
Black Hole is substituted
 by the "stretched horizon" which,
for distant observers, should encode all the microscopical Black
Hole degrees of freedom and, consequently, its observed entropy is
proportional to the area of the horizon consistently
 with the Beckenstein-Hawking law  \footnote{The reason for this is that
 % the effective Black Hole degrees of freedom are in the membrane and
  the entropy must be proportional to the volume of the "space" where the particles live on,
   independently of their nature.}. Our model presents similar behavior if we identify $H$
 with $H_1$ or with some given finite set of H-branes, since they contain all physical degrees
  of
  freedom of the system \footnote{In the membrane paradigm, the stretched horizon will
   disappear for
a freely falling observer which passes the horizon, but to avoid
paradoxes, it is argued a  complementarity principle in which
 the infalling observer cannot report this to a fiducial distant observer \cite{suss,thooft}.
So, the physical meaning of this MGS could also be questioned;
however, the notion of macroscopical observer in itself must be
clarified in this new context, but we will not do it here; so for
simplicity, we have defined the model with respect to distant
fiducial observers.}.

 Note that one could account for many thermodynamical properties of
 Black Holes by assuming an appropriate dynamics for the H-branes.
  For instance, it has been argued that the microscopical degrees of
   freedom of the stretched horizon may be effectively described
 by a Quasi-particles system \cite{K,suss}, however the interpretation of
  this correspondence is obscure. The present approach should help to
 understand it correctly in terms of the microscopical spacetime dynamics.

 In a forthcoming paper \cite{otro}, we construct the simplest example of dynamics consistent with (I),
  by giving an effective action
 for the component subsystems (H-branes and links) which must tend to the four-dimensional Einstein theory
 in the macroscopical limit. This could be helpful to quantize the
 theory.

\vspace{0.4cm}

 {\bf Acknowledgements}:  Special
Thanks are due to R. Scherer and A.L.M.A Nogueira. The author is
supported by CLAF/CNPq.

\end{document}